\documentclass[journal]{IEEEtran}

\usepackage{booktabs}
\usepackage{tabularx}
\usepackage{stfloats}
\usepackage{siunitx}
\normalsize
\usepackage{setspace} 
\usepackage{mathtools}
\usepackage{color}
\usepackage{array}
\usepackage[version=4]{mhchem}
\usepackage{float}
\usepackage{stmaryrd}
\usepackage{soul}
\usepackage{stackrel}
\usepackage{epstopdf}
\usepackage{url}
\usepackage{cite}
\usepackage{tikz}
\usetikzlibrary{automata,arrows,positioning,calc}
\usepackage{bigints}
\usepackage{bbm}
 \usepackage{dsfont}
 \newcolumntype{L}[1]{>{\raggedright\arraybackslash}m{#1}} 
\newcolumntype{C}[1]{>{\centering\arraybackslash}m{#1}}  
\ifCLASSINFOpdf
\else
\fi

\begin{document}

\title{MC for Agriculture: A Framework for Nature-inspired Sustainable Pest Control}
\author{
    Fardad Vakilipoor\IEEEauthorrefmark{1},
    Nora Hirschmann\IEEEauthorrefmark{1},
    Julian Schladt\IEEEauthorrefmark{2},
    Stefan Schwab\IEEEauthorrefmark{1},\\
    Annette Reineke\IEEEauthorrefmark{2},
    Robert Schober\IEEEauthorrefmark{1},
    Kathrin Castiglione\IEEEauthorrefmark{1},
    Maximilian Schäfer\IEEEauthorrefmark{1}
    \\
    \IEEEauthorrefmark{1}Friedrich-Alexander-Universität Erlangen-Nürnberg, Erlangen, Germany\\
    \IEEEauthorrefmark{2}Hochschule Geisenheim University, Geisenheim, Germany
}
%
%

\markboth{MC for Agriculture: A Framework for Nature-inspired Sustainable Pest Control}%
{Vakilipoor \MakeLowercase{\emph{et al.}}}

\maketitle

\begin{abstract}
In agriculture, molecular communication (MC) is envisioned as a framework to address critical challenges such as smart pest control. While conventional approaches mostly rely on synthetic plant protection products, posing high risks for the environment, harnessing plant signaling processes can lead to innovative approaches for nature-inspired sustainable pest control. 
In this paper, we investigate an approach for sustainable pest control and reveal how the MC paradigm can be employed for analysis and optimization. 
In particular, we consider a system where herbivore-induced plant volatiles (HIPVs), specifically methyl salicylate (MeSA), is encapsulated into microspheres deployed on deployed on plant leaves. The controlled release of MeSA from the microspheres, acting as transmitters (TXs), supports pest deterrence and antagonist attraction, providing an eco-friendly alternative to synthetic plant protection products. 
Based on experimental data, we investigate the MeSA release kinetics and obtain an analytical model. To describe the propagation of MeSA in farming environments, we employ a three dimensional (3D) advection-diffusion model, incorporating realistic wind fields which are predominantly affecting particle propagation, and solve it by a finite difference method (FDM).
The proposed model is used to investigate the MeSA distribution for different TX arrangements, representing different practical microsphere deployment strategies. 
Moreover, we introduce the coverage effectiveness index (CEI) as a novel metric to quantify the environmental coverage of MeSA. This analysis offers valuable guidance for the practical development of microspheres and their deployment aimed at enhancing coverage and, consequently, the attraction of antagonistic insects.
\end{abstract}
\IEEEpeerreviewmaketitle
\section{Introduction}
Molecular communication (MC) is a bio-inspired communication paradigm where information is encoded into the properties of molecules \cite{Akyildiz2008}. The envisioned applications of MC range across different scales and domains, including nanotechnology~\cite{Nakano2013}, biomedicine~\cite{9488662}, cellular biology, and smart farming \cite{Aktas_odor}. Therefore, MC systems are often categorized by the application range and environment \cite{expMC_Seb}. While MC systems for medical applications mostly operate in a range of  nano- or micrometers in liquid environments such as the human cardiovascular system \cite{Felicetti_MC}, MC systems for environmental applications operate in a range of meters in large air-based environments \cite{Aktas_odor,vahid_Odor}.

\begin{figure}
    \centering
    \includegraphics[width=0.85\linewidth]{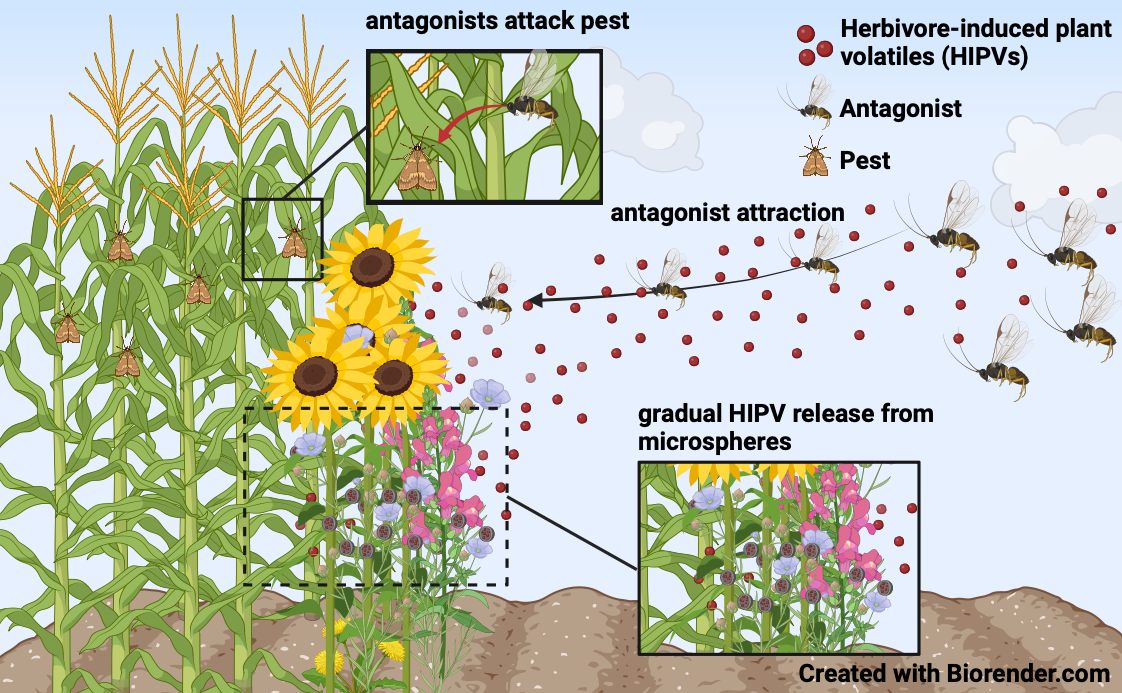}
    \caption{\small Schematic overview of the considered nature-inspired pest-control approach. Microspheres deployed on plant leaves act as TXs of HIPVs to attract antagonists in the environment.}
    \label{fig:Story}
\end{figure}

For environmental and agricultural applications, MC is envisioned to have transformative potential. By harnessing natural plant signaling, MC is envisioned to address critical challenges such as securing global food demands, in particular in the light of preventing crop losses due to pest and diseases. Conventional pest control approaches heavily rely on synthetic plant protection products, which also pose high risks, including soil and water contamination, pollinator decline, pesticide resistance, and threats to human health~\cite{Gould_WE,Goulson_OV,Stamati_Chem}. These drawbacks of chemical pesticides motivate sustainable alternatives which are closely aligned with ecological principles. 
Plants naturally communicate through volatile organic compounds (VOCs), such as herbivore-induced plant volatiles (HIPVs), to coordinate responses to stressors~\cite{Baldwin2006,Aktas_odor}. HIPVs, such as methyl salicylate (MeSA), repel pests and attract antagonists to attack vermin, offering a natural alternative to conventional pest control approaches~\cite{Heil2010,Turlings2018} (see Fig.~\ref{fig:Story}). 
In this paper, we investigate an alternative approach to conventional pest control. In particular, we propose a synthetic MC system which harnesses and mimics natural plant communication processes for the attraction of antagonists. 

Existing works in MC targeting application in agriculture mainly focus on intercellular plant communication and airborne plant communication. For example, the authors of \cite{Hamdan_InCel} investigated how plants increase photosynthesis through intracellular signaling. Similarly, the authors of \cite{Hamdan_ComPlant} demonstrated that multiple action potentials enhance information propagation in plant cell populations, while the work in \cite{Hamdan_InfTheo} compared electrochemical and mechanosensitive signals for intercellular communication in plants. In \cite{Imen_Plant}, machine learning was used to detect water stress in plants via bioelectrical signals, and the authors of \cite{Hughes_plant} modeled plant cell communication as a narrow escape problem. The authors of \cite{Unluturk_ETE} proposed a theoretical end-to-end MC system model for plant pheromone communication, incorporating emission, wind-driven propagation, and reception processes. In \cite{Bilgen:fungi:2024}, the collective communication behavior of plants was investigated within the internet of plants. 
However, most of the previously discussed works remain theoretical and none of them considers the development of practically realizable MC-based concepts for smart pest control. 
In this paper, we propose a practical synthetic MC system as the basis for nature-inspired pest control in agricultural environments. The MC system consists of biodegradable microspheres that act as transmitters (TXs), releasing HIPVs, i.e., MeSA into the environment. The microspheres are deployed on the leaves of plants in flower patches, located next to the crop plants in the field (see Fig.~\ref{fig:Story}). Placing flower patches within fields serve multiple ecological and practical purposes, especially in sustainable or organic farming systems~\cite{su17052018}.
The released MeSA molecules shall attract antagonists such as wasps to attack pests, infesting the crop plants.
This paper focuses primarily on the experimental and theoretical modeling of the MeSA release process as well as on the modeling of MeSA propagation in the environment, while the MeSA-antagonist interaction is left for future work.  
The main contributions of this paper can be summarized as follows:

\begin{itemize}
    \item We propose a novel MC-based approach for agricultural applications, using HIPVs to enhance communication processes between plants, pests, and antagonists, which is envisioned to enable nature-inspired and eco-friendly pest management. 
    \item We experimentally design biodegradable microspheres  acting as TXs for the controlled release of MeSA molecules into the environment, and characterize the experimentally measured release process by the Korsmeyer-Peppas model~\cite{KORSMEYER1981211}.
    \item We develop a three dimensional (3D) partial differential equation (PDE) for MeSA propagation in the environment considering the effects of diffusion, advection under realistic wind fields, and ground absorption, and propose a solution based on the finite difference method (FDM).
    \item We investigate the spatial and temporal MeSA distribution in a large field, considering four different TX topologies, reflecting different practical deployment strategies for microspheres within a flower patch next to crop plants. 
    \item We propose a new metric, the Coverage Effectiveness Index (CEI), which quantifies the fraction of a given volume in which the MeSA concentration exceeds a certain threshold. This index serves as a tool for evaluating and optimizing the effectiveness of microspheres deployment in the field.
\end{itemize}
The remainder of the paper is organized as follows. In Section~\ref{sec:agriMC}, we provide some background on agricultural MC applications, before we introduce the considered HIPV-based MC system and describe the experimental TX design in Section~\ref{sec:hipvMC}. In Section~\ref{sec:sec:model}, we present the mathematical channel model for HIPV propagation in the considered environment, and in Section~\ref{sec:simulation}, we present simulation results for different TX arrangements. Finally, Section~\ref{sec:conclusion} concludes the paper.

\section{Agricultural MC Systems}
\label{sec:agriMC}

Before we describe the considered HIPV-based MC system in Section~\ref{sec:hipvMC}, we provide a short overview on agricultural applications of MC, as MC can be exploited to harness natural plant signaling processes to support crop health and ultimately ecosystem balance. 

Agricultural fields are complex ecosystems where different entities such as plants, microbes, and animals interact through physical, chemical, and biological signals. These interactions between different biological entities, i.e., TXs and receivers (RXs), over different types of channels is facilitated by MC. In the following, we briefly discuss the components of agricultural MC systems. 

In agricultural MC systems, \textbf{TXs} may include plants, microbes, fungi, or insects that emit molecular signals to transmit information \cite{Aktas_odor}.
Plants under herbivore attack emit HIPVs, such as MeSA, to signal distress and repel pests~\cite{Heil2010}. Soil microbes, like the \textit{Bacillus} species, release VOCs that promote plant growth or inhibit pathogens \cite{Ryu2003}. Mycorrhizal fungi exude signaling molecules, such as lipochitooligosaccharides, to establish symbiotic relationships with plant roots~\cite{Olde2014}. Insects, such as aphids, secrete alarm pheromones like (E)-$\beta$-farnesene to warn conspecifics of danger~\cite{Dewhirst2010}. 
    
\textbf{RXs} may include, for example, plants, pests, and antagonists that detect and decode molecular signals, thereby triggering behavioral or physiological responses. Plants, such as maize, perceive HIPVs from attacked neighbors, upregulating defensive genes to enhance resistance~\cite{Baldwin2006}. Herbivorous pests, like \textit{Aphis gossypii} (cotton aphid), are repelled by high concentrations of HIPV, altering their feeding behavior. Antagonist insects, including \textit{Trichogramma} wasps, sense HIPVs to locate herbivore prey, aiding pest control~\cite{Turlings2018}.
    
The \textbf{channel} in agricultural MC systems, over which TXs and RXs communicate is typically air or soil. Environmental factors, including wind, temperature, humidity, and soil composition, influence molecule diffusion, advection, and degradation. For instance, wind-driven advection in air channels transports HIPVs across fields (see Fig.~\ref{fig:Story}), while soil channels rely on slower diffusion.

These components, i.e., TXs, channels, and RXs, facilitate diverse forms of communication within the agricultural ecosystem. In Table~\ref{tab:MC_Eg}, we summarize several examples of known MC mechanisms in such environments.
\begin{table*}[ht]
\centering
\caption{Examples of MC in agriculture ecosystems.}
\label{tab:MC_Eg}
\begin{tabular}{|L{6.1cm}|C{1cm}|C{1cm}|C{1.3cm}|L{6.5cm}|}
  \hline
  \multicolumn{1}{|c|}{\textbf{Mechanism}} & 
  \textbf{TX} & 
  \textbf{RX} & 
  \textbf{Channel} & 
  \multicolumn{1}{c|}{\textbf{Example}} \\
  \hline
  Stressed plants release HIPVs to warn neighboring plants of herbivore attacks, prompting defensive responses such as increased volatile production or toxin synthesis. &
  Plant &
  Plant &
  Air &
  Tomato plants attacked by caterpillars release MeSA, prompting nearby tomato plants to produce proteinase inhibitors to deter feeding~\cite{Rowen2017}. \\
  \hline
  Soil microbes release VOCs that enhance plant growth or immunity. &
  Microbe &
  Plant &
  Soil &
  \textit{Bacillus subtilis} in the soil releases 2,3-butanediol, stimulating maize roots to grow faster and resist fungal pathogens~\cite{Yi2016}.\\
  \hline
   Plants release repellent molecules to deter herbivorous pests. &
   Plant &
   Pest &
   Air &
   Soybean plants emit MeSA when attacked, repelling \textit{Aphis glycines} (soybean aphids) from feeding on their leaves~\cite{chromatography}.\\
   \hline
   Plants emit attractant VOCs to draw antagonist insects that control pest populations. &
   Plant &
   Insect &
   Air &
   Maize plants damaged by caterpillars release linalool, attracting \textit{Trichogramma} wasps to parasitize caterpillar eggs~\cite{Fontana2011}.\\
    \hline
    Insects use pheromones to communicate mating or alarm signals. &
    Insect &
    Insect &
    Air &
     \textit{Aphis gossypii} (cotton aphids) release (E)-$\beta$-farne\-sene when threatened by lady beetles, signaling other aphids to disperse from plants~\cite{Jianwei}.\\
     \hline
     Mycorrhizal fungi release molecules to initiate symbiotic relationships with plant roots, enhancing nutrient uptake. &
     Fungus &
     Plant &
     Soil &
     \textit{Rhizophagus irregularis} releases lipochitooligosaccharides, stimulating \textit{Medicago truncatula} roots to form mycorrhizal connections for enhanced phosphorus uptake~\cite{Maillet2011}.\\
     \hline     
\end{tabular}
\end{table*}
%

\section{HIPV-Based MC}
\label{sec:hipvMC}
In this section, we describe the envisioned MC system and the role of HIPVs in farming environments. Then, we present the experimental design of the proposed microsphere-based TXs and a characterization of the embedded HIPVs, i.e., MeSA, and their release profile.

\begin{figure}[t]
    \centering
    \includegraphics[width=0.9\linewidth]{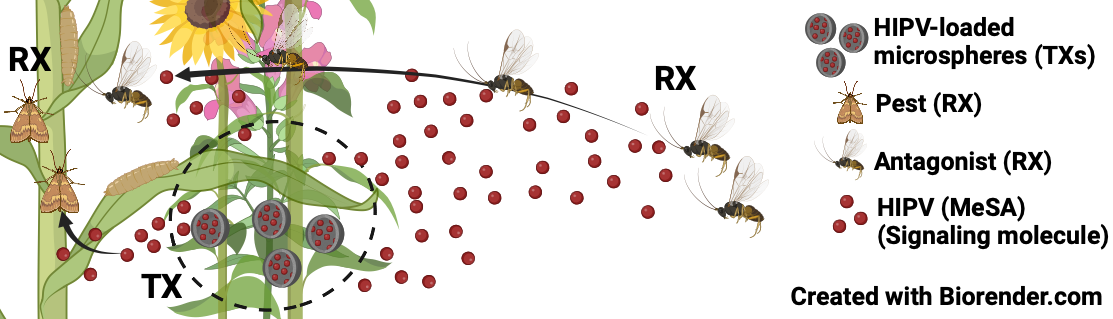}
    \caption{\small Detailed schematic of the proposed HIPV-based MC system. HIPV-loaded microspheres (TXs) release their cargo into the environment to repel pests and to attract antagonists (RXs).}
    \label{fig:story_reduced}
\end{figure}

\subsection{HIPVs}
In this paper, we focus on the scenario shown in Figs.~\ref{fig:Story} and \ref{fig:story_reduced}, where HIPV-loaded microspheres, placed on plant leaves, serve as TXs to attract antagonists. The matrix-based microspheres are composed of biodegradable hydrogenated sunflower oil and encapsulate MeSA, enabling controlled release of MeSA by diffusion and degradation. The HIPVs act as signaling molecules, which influence the behavior of different RXs in the environment, including herbivorous pests (repelled by high MeSA concentrations) and beneficial antagonists like parasitic wasps (attracted to MeSA as cues for prey location).

As shown in Fig.~\ref{fig:story_reduced}, the primary RXs of the released molecules are (i)  \textbf{herbivorous pests}, i.e., aphids, caterpillars, or beetles, which are repelled by MeSA, reducing their feeding activity on crops~\cite{chromatography}, or (ii) \textbf{parasitic wasps}, i.e., parasitoids such as Trichogramma species, Braconid wasps, and members of the Trichogrammatidae and Braconidae families, which are attracted to MeSA, thereby enhancing biological pest control by targeting pest eggs or larvae~\cite{Fontana2011}.

\subsection{Experimental TX Design}
Matrix-based microspheres containing MeSA, acting as TXs, were produced using a spray congealing process. The matrix material, hydrogenated sunflower oil with a melting point of 70~$^\circ$C, was first melted at a controlled temperature above its melting point. MeSA was then homogeneously incorporated into the molten matrix material. The resulting formulation was atomized into a cooled spray tower, where the droplets solidified into microspheres upon cooling. The solidified microspheres were subsequently separated and collected. Particle size analysis was conducted using a Mastersizer $3000$ (Malvern Panalytical, UK) equipped with a dry powder dispersion unit (Aero S). 
The particle size distribution was measured as $\mathrm{Dv}(50) = 180~\si{\micro\meter}$, where $\mathrm{Dv}(50)$ indicates that $50\si{\percent}$ of the particles (by volume) have a diameter below $180~\si{\micro\meter}$. The physical and chemical properties of the used MeSA molecules are summarized in Table~\ref{tab:MeSA_properties}. Each produced microsphere contains $10\si{\percent}$ load of MeSA molecules.
\begin{table}[ht]
\centering
\caption{Physical and chemical properties of MeSA.}
\label{tab:MeSA_properties}
\begin{tabular}{|lc|}
\hline
Molar Mass & $152.149~\si{\gram\per\mole}$ \\ 
Density & $1.174~\si{\gram\per\cubic\centi\metre}$ \\
Melting Point & $-8.6~\si{\degreeCelsius}$ \\
Boiling Point & $222~\si{\degreeCelsius}$ \\
Vapor Pressure & $13~\si{\pascal}$ at $\SI{20}{\degreeCelsius}$ \\
\hline
\end{tabular}
\end{table}

\subsection{MeSA Release Kinetics}
\begin{figure}[t]
    \centering
    \includegraphics[width=1\linewidth]{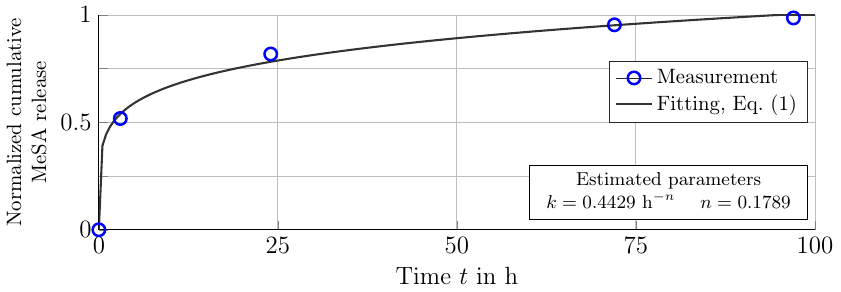}
    \caption{Normalized cumulative release of MeSA from microspheres over time, showing experimental data and the fitted Korsmeyer-Peppas model with $k = 0.4429 \, \text{h}^{-n}$ and $n = 0.1789$.}
    \label{fig:release_curve}
\end{figure}

To incorporate the proposed microsphere-based TXs into the channel model in Section~\ref{sec:sec:model}, we investigate their MeSA release kinetics in the following. To measure the release kinetics experimentally, the microspheres were exposed to air at room temperature. Samples were prepared every few hours by adding $0.05~\si{\gram}$ of MeSA microspheres and $1~\si{\micro\liter}$ of an internal standard (IS) to vials. The amount of residual MeSA in the samples was detected using head space solid-phase microextraction (HS-SPME) and IS calibration. Therefore, an SPME fiber was exposed to the samples, allowing the MeSA and the IS to partition between the fiber and the sample until an equilibrium was reached. Then, the SPME fiber was analyzed using gas chromatography with flame ionization detection. By calculating the residual MeSA concentration in the microspheres at different times, the normalized cumulative release kinetics of MeSA over time were obtained (see Fig.~\ref{fig:release_curve}). 

To analyze the measured release kinetics, we fitted the experimental data to the Korsmeyer-Peppas model~\cite{KORSMEYER1981211}, a semi-empirical model that is commonly used to describe the diffusion-controlled release of molecules from polymeric matrix systems. The model describes the cumulative amount of released molecules $M(t)$ as follows
\begin{align}\label{eq:peppas}
    \frac{M(t)}{M_\infty} = k t^n\,,
\end{align}
where $M_\infty$ is the total amount of molecules released at infinite time (equivalent to the initial loading if everything is released), $k$ is the release rate constant, and $n$ is the release exponent indicative of the release mechanism~\cite{KORSMEYER1981211}. Figure~\ref{fig:release_curve} shows the measured release kinetics (blue markers), the results from fitting \eqref{eq:peppas} to the measurement data (black curve), and the model parameters $k$ and $n$ estimated during the fitting process. From the figure it can be observed that the experimentally measured release kinetics of the considered microshpere-based TX are well\footnote{The fitting quality is quantified by the coefficient of determination ($R^2 = 0.9969$). The 95\% confidence intervals are $k \in [0.3647, 0.5273]$ and $n \in [0.1343, 0.2273]$.} approximated using \eqref{eq:peppas}. 
%

%
\section{MC Channel Model}\label{sec:sec:model}
\begin{figure}[t]
    \centering
    \includegraphics[width=0.85\linewidth]{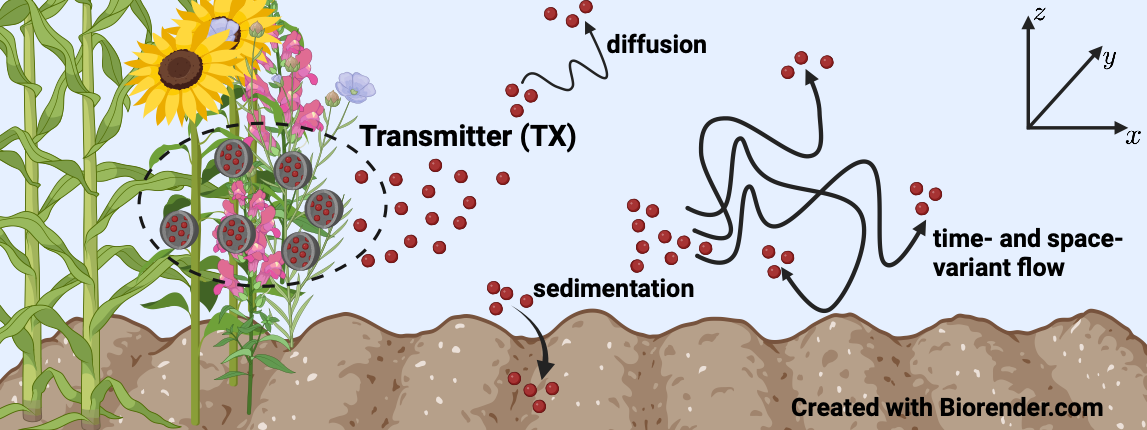}
    \caption{\small Schematic of MeSA propagation and environmental effects in a field after release from the TXs located in the flower patch.}
    \label{fig:hipvprop}
\end{figure}

As previously mentioned, in this paper, we mainly focus on the release of MeSA from microspheres, acting as TXs, and their propagation in the environment (see Fig.~\ref{fig:hipvprop}), while the interaction of MeSA with antagonists and pests is left for future work. Therefore, in this section, we first introduce a mathematical model for the propagation of MeSA, with a focus on the wind model. Then, we extend the TX release kinetics, obtained in Section~\ref{sec:hipvMC} to multiple microspheres and incorporate it into the mathematical model. Finally, we propose a new metric, the CEI that quantifies the effective coverage of MeSA in the environment.

\subsection{Physical Channel}
To describe MeSA propagation in the environment, we employ a 3D advection-diffusion PDE for the MeSA concentration $C(x, y, z, t)$, in $\text{molecules}~\si{\per\cubic\meter}$, over space $(x, y, z)$ and time $t$. The PDE accounts for diffusion, wind-driven advection, and includes a source term representing the MeSA release from multiple TXs, as follows
\begin{align}\label{eq:PDE}
    \frac{\partial C}{\partial t} = D \nabla^2 C - \mathbf{v}(x, y, z, t) \cdot \nabla C + S(x, y, z, t)\,.
\end{align}
Here, $\nabla^2$, $\nabla$ and $\cdot$ are the Laplacian operator, the gradient operator with respect to space, and the inner product, respectively. Variable $D$ denotes the diffusion coefficient in $\si{\square\meter\per\second}$, $\mathbf{v}(x, y, z, t)$ is the time- and space-variant wind velocity in $\si{\meter\per\second}$, and $S(x, y, z, t)$ is the source term in molecules $\si{\per\cubic\meter\per\second}$. 
The initial condition for \eqref{eq:PDE} is $C(x, y, z, 0) = 0$. The sedimentation of MeSA at the ground is modeled by a fully absorbing boundary at $z = 0$, i.e., $C(x, y, 0, t) = 0$.

\subsection{Wind Field Model}
The dominant propagation mechanism of molecules in the considered air-based agricultural scenario, is the wind. Therefore, to model a realistic environment, we consider a time- and space-dependent wind velocity in \eqref{eq:PDE}. The wind is defined over the field with a mean speed of $\bar{v} = 0.5 \,\si{\meter\per\second}$, indicating a light air~\cite[Table 5.1]{wmo2018guide}. Wind varies diurnally in speed and direction. Moreover, we considered stochastic terms that contribute to wind direction and speed in every time instant and each point in space (see Appendix~\ref{ap:wind_model}). To account for the shear near the ground ($z = 0$), we adopt a logarithmic vertical profile for the wind component in vertical direction~\cite{Stull1988}.

\subsection{TX Model}
The source term $S(x, y, z, t)$ in \eqref{eq:PDE} models the MeSA release from $N$ microspheres, equivalent to multiple TXs distributed in space. For $N$ microspheres, the source term follows as
\begin{align}\label{eq:source}
    S(x, y, z, t) = \sum_{l=1}^N q(t) \delta(x - x_l) \delta(y - y_l) \delta(z - z_l)\,,
\end{align}
where $\delta(\cdot)$ and $(x_l,y_l,z_l)$ are the Dirac delta function and the position of the $l$-th microsphere, respectively. The release rate, $q(t) = \frac{\mathrm{d} M(t)}{\mathrm{d}t}$, is the derivative of the cumulative released MeSA in \eqref{eq:peppas}. This formulation of the source term in \eqref{eq:source} incorporates the experimentally obtained release kinetics into PDE \eqref{eq:PDE}, enabling the accurate modeling of MeSA release from multiple TXs.

\subsection{CEI}
The goal of the proposed MC system is to efficiently distribute MeSA molecules in the environment, and ultimately to attract antagonists. Therefore, to asses the performance of the system, we introduce a new measure, the CEI, to quantify the effective MeSA coverage in the environment. The CEI indicates the time-averaged proportion of a considered volume $\mathrm{V}$ in which the concentration of MeSA exceeds a threshold $C_{\mathrm{th}}$.
The CEI is defined as follows
\begin{align}
\text{CEI}(t,C_{\text{th}}) &= \frac{1}{t} \int_{0}^{t} \frac{1}{\mathrm{V}}\int_{\mathrm{V}} \mathds{1}(C(x,y,z,\tau) \geq C_{\mathrm{th}}) \, \mathrm{d}\mathrm{V} \, \mathrm{d}\tau\,,\label{eq:CEI}
\end{align}
where $\mathds{1}(\cdot)$ is the indicator function, and $t$ is the total time interval considered. The volume $\mathrm{V}$ is the volume over which MeSA coverage is assessed, which can be either the entire environment, or a sub-volume of interest where the coverage should be measured. The CEI quantifies the spatiotemporal fraction of $\mathrm{V}$ where the MeSA concentration exceeds a threshold value $C_{\mathrm{th}}$, and its values can vary between 0 and 1, where $1$ corresponds to the maximum effective coverage. Threshold $C_{\mathrm{th}}$ can be chosen as, e.g., the minimum MeSA sensing threshold of antagonists in the environment or the repelling threshold of herbivorous pests. To this end, the CEI can be employed to asses how efficiently MeSA is distributed in the environment, and how this efficiency is influenced by parameters such as different TX arrangements or microsphere compositions. 
%

\section{Simulation Results}\label{sec:simulation}
In the following, we simulate the propagation of MeSA released from microsphere-based TXs into the environment. The simulation is based on the system model proposed in Section~\ref{sec:sec:model}. First, we describe the numerical simulation approach used for solving \eqref{eq:PDE}, then we describe the scenarios considered for evaluation. Finally, we analyze the MeSA concentration profiles and CEI for different TX arrangements. 

\subsection{FDM and Stability}\label{sec:fdm}
PDE \eqref{eq:PDE} is solved using FDM~\cite{FDM_Book} by discretizing the spatial and temporal domains into a grid with spatial steps $\Delta x = \Delta y = 5 \, \text{m}$, $\Delta z = 0.5 \, \text{m}$, and time step $\Delta t = 2 \, \text{s}$. The grid comprises $41 \times 41 \times 11$ points ($x \in [-100, 100]~\text{m}$, $y \in [-100, 100]~\text{m}$, $z \in [0, 5]~\text{m}$) over $43200$ time steps (24 hours). 
The diffusion term in \eqref{eq:PDE} is approximated using central differences as follows
\begin{align}
\begin{split}
    D \nabla^2 C \approx D \Bigg(& \frac{C_{i+1,j,k} - 2C_{i,j,k} + C_{i-1,j,k}}{\Delta x^2} 
    +\\& \frac{C_{i,j+1,k} - 2C_{i,j,k} + C_{i,j-1,k}}{\Delta y^2} +\\&\frac{C_{i,j,k+1} - 2C_{i,j,k} + C_{i,j,k-1}}{\Delta z^2} \Bigg)\,,
    \end{split}
    \label{eq:fdm:diff}
\end{align}
where $ C_{i,j,k} $ is the concentration at grid point $ (x_i, y_j, z_k) $ with $ i $, $ j $, and $ k $ indexing the discretized $ x $-, $ y $-, and $ z $-axes, respectively. Each term in \eqref{eq:fdm:diff} approximates the second partial derivative in one spatial direction, e.g., $ \frac{\partial^2 C}{\partial x^2} $ for the $x$-direction, using central differences, evaluates the concentration at neighboring grid points.

\begin{figure}[t]
    \centering
    \includegraphics[width=\linewidth]{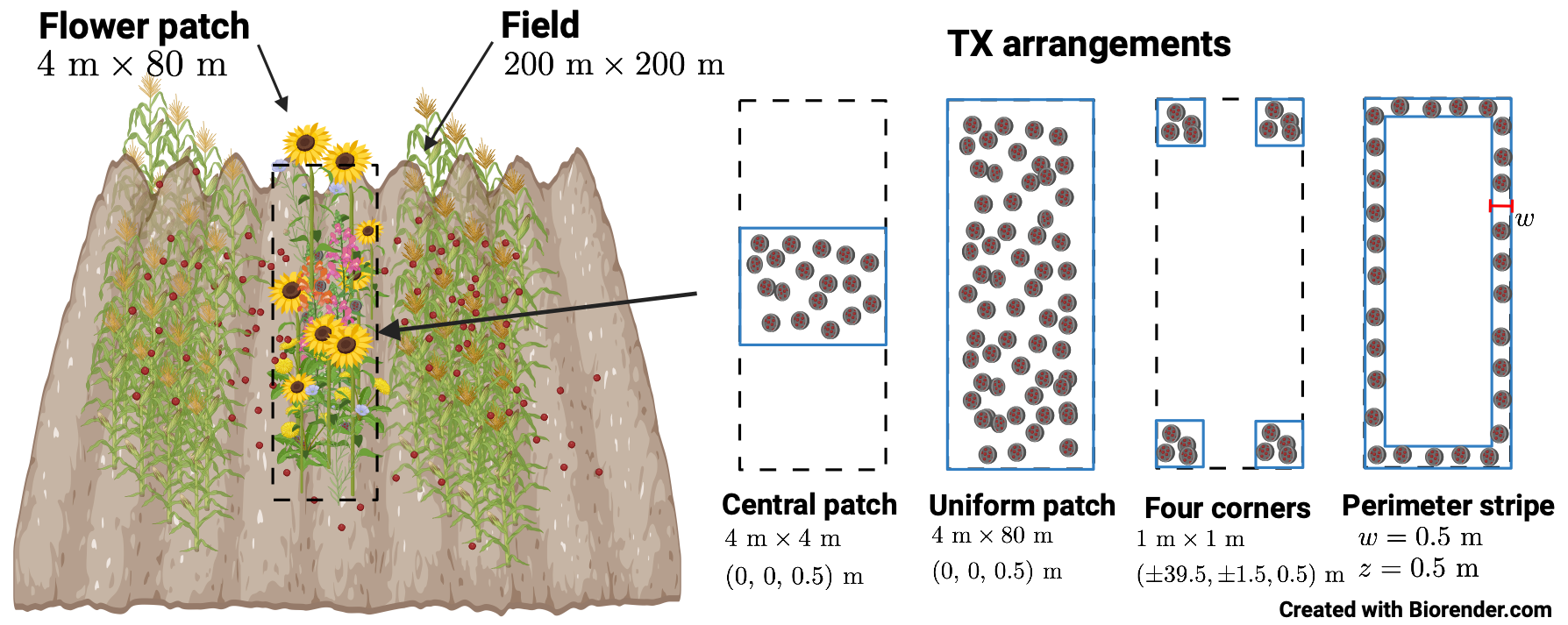}
    \caption{\small Simulation scenario (left) and different TX arrangements and microsphere distributions.}
    \label{fig:sim_scenario}
\end{figure}

The advection term in \eqref{eq:PDE},
representing the MeSA propagation due to velocity $ \mathbf{v} = [v_x, v_y, v_z] $, is computed using a first-order FDM. For each grid point $ (x_i, y_j, z_k) $, the advection term is approximated as $ -\left( v_x \frac{\partial C}{\partial x} + v_y \frac{\partial C}{\partial y} + v_z \frac{\partial C}{\partial z} \right) $, where the spatial derivatives are computed using an upwind scheme that adapts to the sign of the velocity components. Specifically, for a velocity component $ v_x \geq 0 $, the derivative is approximated as $ \frac{\partial C}{\partial x} \approx \frac{C_{i,j,k} - C_{i-1,j,k}}{\Delta x} $, using the upstream point $ (x_{i-1}, y_j, z_k) $; for $ v_x < 0 $, it is $ \frac{\partial C}{\partial x} \approx \frac{C_{i+1,j,k} - C_{i,j,k}}{\Delta x} $, using $ (x_{i+1}, y_j, z_k) $. Similar approximations are applied for the $ y $- and $ z $-components $ v_y $ and $ v_z $, respectively.

\begin{figure*}[ht]
    \centering
    \begin{minipage}{0.33\linewidth}
        \includegraphics[width=0.7\linewidth]{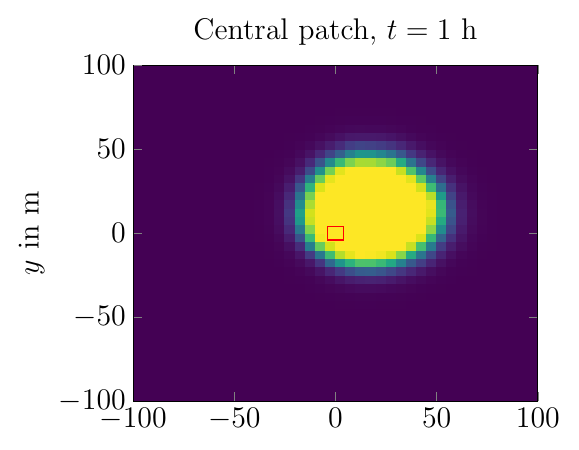}
    \end{minipage}\hfill
    \begin{minipage}{0.33\linewidth}
        \includegraphics[width=0.7\linewidth]{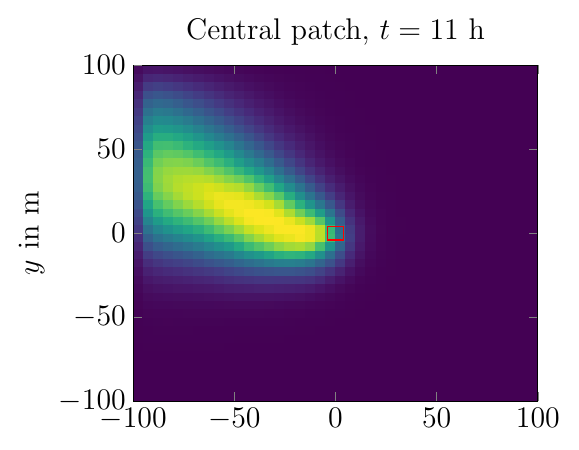}
    \end{minipage}
    \begin{minipage}{0.33\linewidth}
        \includegraphics[width=0.8\linewidth]{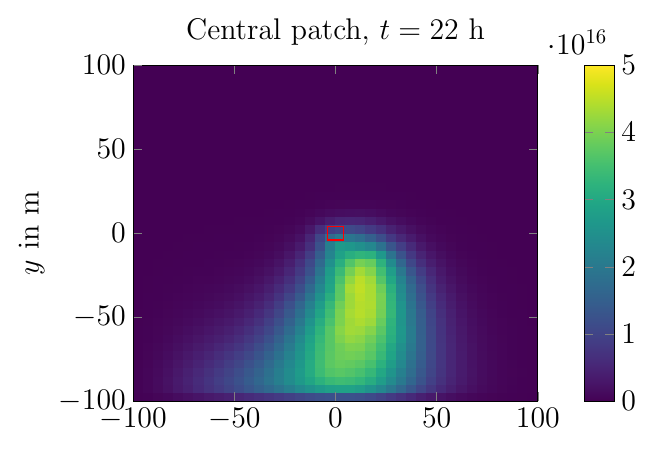}
    \end{minipage}\\[-0.8em]
    \begin{minipage}{0.33\linewidth}
        \includegraphics[width=0.7\linewidth]{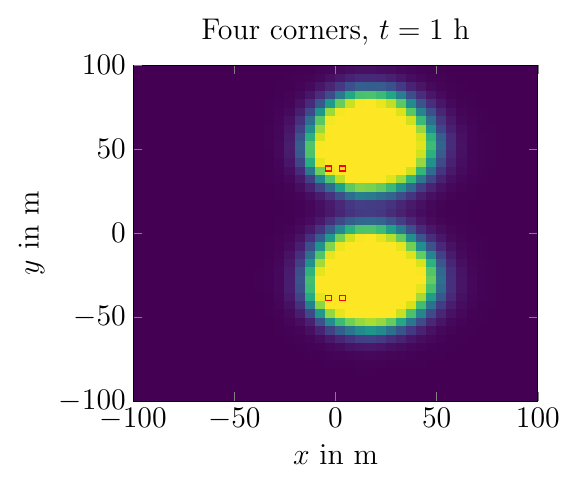}
    \end{minipage}\hfill
    \begin{minipage}{0.33\linewidth}
        \includegraphics[width=0.7\linewidth]{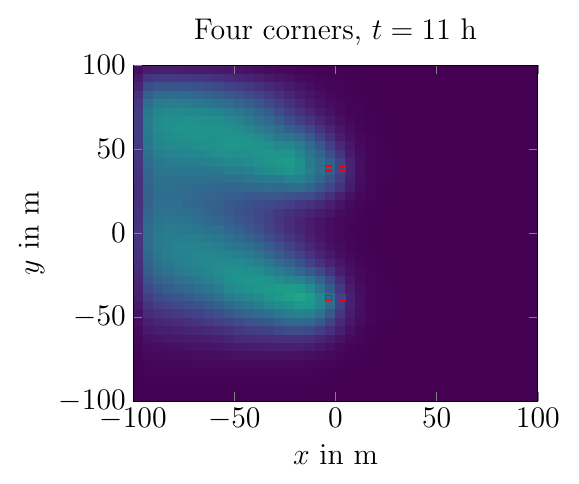}
    \end{minipage}
    \begin{minipage}{0.33\linewidth}
        \includegraphics[width=0.8\linewidth]{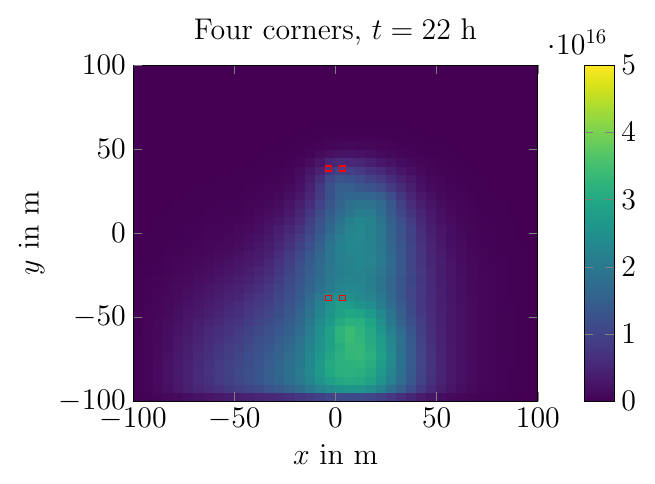}
    \end{minipage}
    \caption{\small Mean MeSA concentration over space, centered at height $z_\mathrm{o} = 2 \, \text{m}$ at times $t \in \{1, 11, 22\} \, \si{\hour}$, for the central patch (top) and four corners (bottom) scenarios. The initial deployment of the microspheres for both scenarios according to Fig.~\ref{fig:sim_scenario} is indicated by red squares. }
    \label{fig:Heat}
\end{figure*}

The source term \eqref{eq:source} is incorporated into the FDM computation as follows. The individual microspheres, with continuous coordinates $(x_l, y_l, z_l)$, are mapped to the nearest grid point $ (x_i, y_j, z_k) $ by minimizing the absolute distance in each spatial dimension. 
The contributions of all microspheres, mapped to the same grid point are summed up, and the resulting source term is added to the concentration update in each time step using a forward Euler scheme, i.e., $ C_{i,j,k}^{n+1} = C_{i,j,k}^n + \Delta t (D \nabla^2 C - \mathbf{v} \cdot \nabla C + S_{i,j,k}) $. 

The boundary conditions are enforced as $C = 0$ at $z = 0$, modeling the sedimentation at the ground of the field. The boundary conditions at the lateral ($x = \pm 100 \, \text{m}, y = \pm 100 \, \text{m}$) and top ($z = 5 \, \text{m}$) boundaries are the first-order outflow conditions in order to model an open space where MeSA can leave the domain. For a boundary with outward normal, the outflow condition is based on the advection equation $\frac{\partial u}{\partial t} + \mathbf{v} \cdot \nabla C = 0$. This implies that molecules are carried out by the wind without additional diffusion at the boundary. In discrete form, for example, at $x = 100$ (index $i = n_x$), if $v_x > 0$ (outflow), the condition is approximated as $C_{n_x,j,k}^{n+1} = C_{n_x,j,k}^n - \Delta t \cdot v_x \cdot \frac{C_{n_x,j,k}^n - C_{n_x-1,j,k}^n}{\Delta x}$, using an upwind scheme to account for the wind direction. For inflow ($v_x \leq 0$), $C_{n_x,j,k}^{n+1} = 0$, i.e., it is assumed that no external MeSA enters. Similar equations apply at the other boundaries. These outflow conditions allow MeSA to exit the domain naturally, driven by wind, simulating an open field. 
%
The stability of the adopted FDM approach is ensured by satisfying the Courant-Friedrichs-Lewy (CFL) conditions~\cite{Courant1928}. For advection, CFL requires, $
    \Delta t \leq \min \left( \frac{\Delta x}{\max |v_x|}, \frac{\Delta y}{\max |v_y|}, \frac{\Delta z}{\max |v_z|} \right)\,
$,
and for diffusion $
    \Delta t \leq \frac{\min(\Delta x^2, \Delta y^2, \Delta z^2)}{2 D \cdot 3}\,.
$
The chosen $\Delta t = 2 \, \text{s}$ satisfies both conditions.

\subsection{Simulation Scenarios}
In the following, we simulate the propagation of MeSA in a volume of size $200 \, \text{m} \times 200 \, \text{m} \times 5 \, \text{m}$ ($\mathrm{V}_{\mathrm{sim}} = 2\times10^{5}~\si{\cubic\meter}$) over $t_{\mathrm{sim}}=24~\si{\hour}$. As described above, the concentration $C(x,y,z,t)$ of MeSA is obtained by solving PDE \eqref{eq:PDE} by FDM. 
The MeSA diffusion coefficient in air is $D = 1 \times 10^{-5} \, \si{\square\meter\per\second}$ (see Appendix~\ref{ap:diffusion}). We use a realistic wind field (diurnal variations, spatial stochasticity, logarithmic vertical profile), and the source term comprises $N = 7.063
\times 10^{8}$ microspheres, equivalent to $200~\si{\gram}$ of MeSA (see Appendix~\ref{ap:N_HIPVs}). 
For the TX arrangement, we assume a rectangular flower stripe~\cite{su17052018} of $\SI{4}{\meter}$ by $\SI{80}{\meter}$ located in the center of the field $200~\si{\meter}\times200~\si{\meter}$, see Fig.~\ref{fig:sim_scenario}. Within the flower stripe, the microspheres can be distributed, yielding different TX arrangements \footnote{In practice, the microspheres can be distributed on the flower stripe via different techniques such as spraying, dusting, and irrigation to stay on the leaves of the plants. Here, we assume that all microspheres are located at $z = 0.5~\si{\meter}$.}. In particular, we investigate four different TX arrangements to investigate the impact of microsphere distribution on the MeSA dispersion in the field (see Fig.~\ref{fig:sim_scenario}, right hand side):

\begin{itemize}
    \item \textbf{Central patch}: Microspheres are uniformly distributed within a $4 \, \text{m} \times 4 \, \text{m}$ patch centered at $(0, 0, 0.5)~\si{\meter}$. 
    \item \textbf{Uniform patch}: Microspheres are uniformly spread across a $4 \, \text{m} \times 80 \, \text{m}$ patch centered at $(0, 0, 0.5)~\si{\meter}$.
    \item \textbf{Four corners}: Microspheres are distributed equally among four $1 \, \text{m} \times 1 \, \text{m}$ patches at the corners ($(\pm 39.5, \pm 1.5, 0.5)$~\si{\meter}) of the flower stripe, with $N/4$ microspheres per patch.
    \item \textbf{Perimeter stripe}: Microspheres form a $w = 0.5 \, \text{m}$-wide stripe along the perimeter of the flowering stripe at a height of $z = 0.5~\si{\meter}$.
\end{itemize}
In each scenario, we use $N$ microspheres, allowing a comparison of the resulting spatial distribution of MeSA in the environment. 

\subsection{MeSA Concentration Profiles}
In the following, we investigate the concentration of MeSA over $x$ and $y$ centered at height $z_\mathrm{o} = 2~\si{\meter}$, for three different time instances, i.e., $t \in \{1, 11, 22\} \, \si{\hour}$. The concentration profiles were computed by averaging concentrations over the domain $z = [z_\mathrm{o} - 0.5, z_\mathrm{o} + 0.5]$. In Fig.~\ref{fig:Heat}, the concentration profiles for two considered TX arrangements, i.e., \textit{central patch} and \textit{four corners} (see Fig.~\ref{fig:sim_scenario}), are shown.  

\begin{figure}[t]
    \centering
    \includegraphics[width=0.9\linewidth]{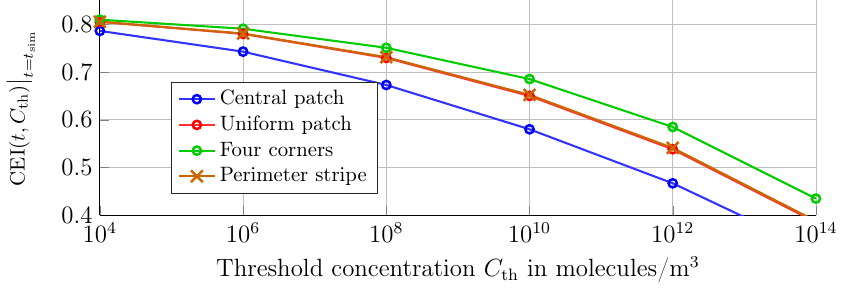}\\
    \includegraphics[width=0.9\linewidth]{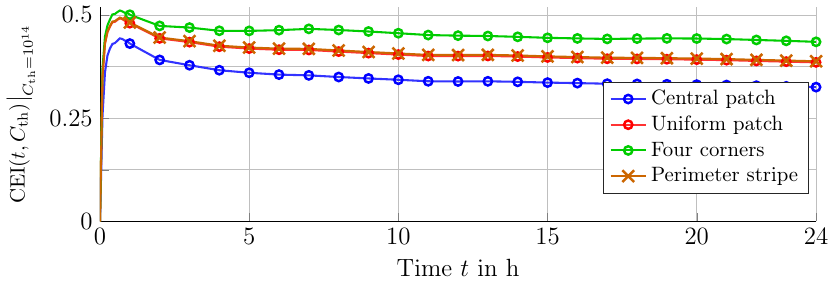}
    \caption{\small Top: Coverage index CEI from \eqref{eq:CEI} versus threshold concentration $C_\mathrm{th}$, averaged over $t_\mathrm{sim} = 24~\si{\hour}$ for the four considered TX arrangements. Bottom: CEI from \eqref{eq:CEI} versus time for a fixed threshold concentration $C_\mathrm{th}=10^{14}$ for all four considered TX arrangements.}
    \label{fig:cei_th}
\end{figure}

We first investigate the influence of the wind profile on the MeSA distribution. From Fig.~\ref{fig:Heat} the effect of the diurnal factor $d(t)$, modeling the daily wind speed variations (see Appendix~\ref{ap:wind_model}), can be seen. At $t = 1~\si{\hour}$ (left hand side), when $d(t)$ is small, the MeSA is only spread over a few meters, for $t = 11~\si{\hour}$ (center plots), the influence of the wind leads to much stronger spreading of MeSA over the field. Furthermore, we can observe how daily wind direction changes (see $\theta$ in Appendix~\ref{ap:wind_model}) influence the main propagation direction of MeSA. Moreover, we can also observe that the different TX arrangements lead to different concentration profiles. For the central patch (top), one clear plume can be observed, while the four corners arrangement (bottom) leads to two plumes, indicating that the latter arrangement leads to a better effective coverage (see also Fig.~\ref{fig:cei_th}). 

\subsection{Effective MeSA Coverage}
In the following, we compare the effective MeSA coverage in the entire volume $\mathrm{V}_{\mathrm{sim}}$ for the different considered TX arrangements. First, we investigate the CEI for different threshold concentrations $C_\mathrm{th}$, which could correspond to the sensitivity of antagonists in future works. $\mathrm{CEI}(t,C_\mathrm{th})\vert_{t = t_\mathrm{sim}}$ is computed according to \eqref{eq:CEI} with $V = V_\mathrm{sim}$ and time $t = t_{\mathrm{sim}}$. Figure~\ref{fig:cei_th} (top) shows the CEI versus $C_{\text{th}}$, ranging from $10^4$ to $10^{14} \, \text{molecules}~\si{\per\cubic\meter}$, for all four considered TX arrangements. 
We can observe that all considered TX arrangements lead to a similar coverage in $\mathrm{V}$ for low values of $C_\mathrm{th}$. However, as $C_\mathrm{th}$ increases, the four corner arrangement yields the highest coverage, while the central patch arrangement leads to the lowest CEI. Interestingly, the uniform and perimeter TX arrangements lead to similar effective coverage in the environment. These observations indicate that the TX arrangement, i.e., the deployment of the microspheres in the flower patch, is a crucial design parameter as it has a strong influence on the overall MeSA coverage in the environment. When the threshold concentration $C_\mathrm{th}$ is related to the sensitivity of antagonists, the CEI can be used to optimize the TX arrangement to maximize the number of attracted antagonists.

Next, in Fig.~\ref{fig:cei_th} (bottom) we investigate the CEI versus time for a fixed threshold concentration $C_\mathrm{th}$. $\mathrm{CEI}(t,C_\mathrm{th})\vert_{C_\mathrm{th} = 10^{14}}$ is calculated according to \eqref{eq:CEI}, with $V = V_\mathrm{sim}$ and $C_\mathrm{th} = 10^{14}$ $\text{molecules}~\si{\per\cubic\meter}$. First, we observe from the bottom plot in Fig.~\ref{fig:cei_th} that the results confirm the observations made from the top plot. In particular, the four corner patch (green) leads to the highest coverage, the uniform (red) and perimeter patch (orange) lead to the same coverage, and the central patch (blue) to the lowest coverage. These results also confirm the observations made from the concentration profiles, where the four corner arrangement (bottom row plots in Fig.~\ref{fig:Heat}) leads to a higher observable coverage compared to the central patch scenario.
\section{Conclusion}\label{sec:conclusion}
In this paper, we proposed an MC system for nature-inspired, sustainable pest control based on HIPV-loaded microspheres for the attraction of antagonists in farming environments. 
Based on experimental results, we characterized the release kinetics of MeSA molecules encapsulated in microspheres, serving as TXs in the considered MC system. Then, we presented a mathematical model for the propagation of MeSA, released from the microspheres, in a farming environment, incorporating various realistic environmental effects including diffusion, advection, and ground absorption. 
To evaluate different practical deployment strategies for the microspheres across agricultural fields, we introduced the CEI, a metric that quantifies the percentage of spatial coverage exceeding a concentration threshold over a specified time duration. Then, we analyzed the MeSA coverage in the environment for four different TX arrangements, representing different practical deployment strategies for the MeSA-loaded microspheres. Our results revealed that the initial deployment has a significant influence on coverage.

This paper lays the foundation for MC in agriculture and reveals its potential for the design of novel pest-control strategies, e.g., to reduce dependence on chemical pesticides and enhance ecological balance. While this was an initial study, there are several topics for future research. First, we will extend the proposed framework by integrating antagonistic agents and investigate their behavioral responses to MeSA gradients.
Moreover, we will investigate the influence of different microsphere formulation and MeSA loadings, yielding different release profiles, on the overall coverage. In future simulations, we will also consider the re-emission of MeSA by plants in response to the initially propagated MeSA.
Furthermore, a practical deployment model tailored for viticulture will be developed, incorporating vineyard land slope. Finally, although we considered realistic wind profiles for the velocity field, the incorporation of real-time wind data\footnote{\url{www.vitimeteo.de}} will enhance the model's applicability for diverse agricultural environments. 
\appendices
\section{Wind Model}
\label{ap:wind_model}
To simulate the transport of MeSA via \eqref{eq:PDE}, a dynamic wind model was developed to capture realistic atmospheric conditions over a 24-hour period. The wind velocity field $\mathbf{v} = (v_{x}, v_{y}, v_{z})$ in \eqref{eq:PDE} is defined on a spatio-temporal grid $({x_{i}},{y_{j}},{z_{k}},t)$ as described in Section~\ref{sec:fdm}. Wind speed and direction include a mean component~\cite{wmo2018guide}, diurnal variation~\cite{ephrath1996modelling}, temporal and spatial noise~\cite{LOUKATOU20181328}, and a vertical shear~\cite{Stull1988} profile to model boundary layer effects near the ground. The mathematical formulations of the different effects were chosen to balance realism and computational simplicity, and are summarized in the following. The horizontal speed of the wind at time $ t $ (in seconds) is given by~\cite[Equation 4]{ephrath1996modelling}
\[
v_{\mathrm{s}}({x_{i}},{y_{j}},{z_{k}},t) = \bar{v} \cdot \left( 1 + 0.5 \cdot d(t) \right) \cdot \left( 1 + \eta_{\text{speed}}({x_{i}},{y_{j}},{z_{k}}, t) \right)\,,
\]
where $ \bar{v} = 0.5 \, \si{\meter\per\second} $ is the mean wind speed, $ d(t) = \sin\left( \frac{2\pi t}{24 \cdot 3600} \right) $ is the diurnal factor, $ \eta_{\text{speed}} \sim \mathcal{N}(0, 0.03) $ is spatial and temporal speed noise following a Gaussian distribution with zero mean and variance $0.03$~\cite[Section~4.2.1]{LOUKATOU20181328}.
The mean wind speed of $ \bar{v} = 0.5 \,\si{\meter\per\second} $, classified as light air on the Beaufort scale, reflects calm conditions, typical for agricultural fields, causing slight leaf movement without significant turbulence~\cite{wmo2018guide}. The diurnal factor $ d(t) $ models daily wind speed variations consistent with solar-driven atmospheric dynamics. 

The wind direction varies during the day and it is defined by angle $\theta$ as follows~\cite[Section~3]{Daikoku2007}
\[
\theta({x_{i}},{y_{j}},{z_{k}},t) = \frac{2\pi t}{24 \cdot 3600} + \eta_{\text{dir}}({x_{i}},{y_{j}},{z_{k}},t)\,,
\]
with $ \eta_{\text{dir}} \sim \mathcal{N}(0, 1.5) $ as spatial and temporal directional noise following a Gaussian distribution with zero mean and variance $1.5$ to mimic shifts~\cite{LOUKATOU20181328}.
The horizontal components of the wind are
\[
v_x({x_{i}},{y_{j}},{z_{k}},t) = v_{\mathrm{s}}({x_{i}},{y_{j}},{z_{k}},t)  \cos(\theta({x_{i}},{y_{j}},{z_{k}},t))\,,
\]
\[ v_y({x_{i}},{y_{j}},{z_{k}},t) = v_{\mathrm{s}}({x_{i}},{y_{j}},{z_{k}},t)  \sin(\theta({x_{i}},{y_{j}},{z_{k}},t))\,.
\]
The vertical wind speed model assumes a logarithmic profile, approximating boundary layer shear where wind speed increases with height, scaled by 10\% of the horizontal speed to reflect weaker vertical mixing near the ground. To this end, the vertical component is defined as~\cite[Section~9.5]{Stull1988}
\[
v_z({x_{i}},{y_{j}},{z_{k}},t) = 0.1 \cdot v_{\mathrm{s}}({x_{i}},{y_{j}},{z_{k}},t) \cdot \ln\left( 1 + \frac{z_k}{z_{\text{ref}}} \right)\,,
\]
where $ z_{\text{ref}} = 5 \, \text{m} $ is the reference height.

\section{MeSA Diffusion Coefficient}\label{ap:diffusion}
The diffusion coefficient of MeSA in air is estimated using the Chapman-Enskog theory for gas diffusion. To this end, the diffusion coefficient $ D $ of MeSA can be obtained as follows~\cite[Equation 17.3-12]{Bird2002}
\[
D = \frac{0.0018583 \mathrm{T}^{3/2} \sqrt{\frac{1}{M_{\mathrm{MeSA}}} + \frac{1}{M_{\mathrm{Air}}}}}{P \sigma_{\mathrm{MeSA,Air}}^2 \Omega_D}\,,
\]
where $ \mathrm{T} = 298 \, \text{K} $ (25°C), $ P = 1 \, \text{atm} $, $ M_{\mathrm{MeSA}} = 152.149 \, \si{\gram\per\mol}$, and $ M_{\mathrm{Air}} = 28.97 \, \si{\gram\per\mol} $, are the temperature, pressure, molar mass of MeSA, and molar mass of air, respectively. 
The average collision diameter $ \sigma_{\mathrm{MeSA,Air}} $ is computed as $ \sigma_{\mathrm{MeSA,Air}} = \frac{\sigma_{\mathrm{MeSA}} + \sigma_{\mathrm{Air}}}{2} $, with $ \sigma_{\mathrm{Air}} = 3.7 \, \text{\AA} $ for air and $ \sigma_{\mathrm{MeSA}} \approx 5.06 \, \text{\AA}$ for MeSA, estimated from the molar volume of MeSA, $ V_{\mathrm{MeSA}} = \frac{M_{\mathrm{MeSA}}}{\rho_{\mathrm{MeSA}}} = \frac{152.149}{1.174} \approx 129.51 \, \si{\cubic\centi\meter\per\mol}$, where $\rho_{\mathrm{MeSA}}$ is the density of MeSA in $\si{\gram\per\cubic\centi\meter}$.
Assuming a collision integral of $ \Omega_D \approx 1.0 $, the diffusion coefficient of MeSA follows as 
$
D \approx 0.101 \, \si{\square\centi\meter\per\second} \approx 1.0 \times 10^{-5} \, \si{\square\meter\per\second}$.

\section{Number of Microspheres}\label{ap:N_HIPVs}
To determine the number of microspheres and MeSA molecules in $2~\si{\kilo\gram}$ of microspheres with $10\si{\percent}$ MeSA loading, we assume spherical microspheres with diameter $ Dv(50) = 180 \, \mu\text{m}$, and a matrix material (hydrogenated sunflower oil) density of $ \rho_{\text{matrix}} = 0.9 \times 10^3 \, \si{\kilo\gram\per\cubic\meter}$ at 25°C. The MeSA density is $ \rho_{\text{MeSA}} = 1.174 \times 10^3 \, \si{\kilo\gram\per\cubic\meter} $, and its molar mass is $ M_A = 152.149 \, \si{\gram\per\mol} $. The microsphere volume is:
\[
V = \frac{4}{3} \pi r^3 = \frac{4}{3} \pi (9.0 \times 10^{-5})^3 \approx 3.054 \times 10^{-12} \, \text{m}^3\,.
\]
The theoretical density of a single microsphere is:
\[
\rho_{\text{eff}} = 0.1 \times 1.174 \times 10^3 + 0.9 \times 0.9 \times 10^3 = 927.4 \, \si{\kilo\gram\per\cubic\meter}\,.
\]
The mass of one microsphere is:
\[
m = \rho_{\text{eff}} \cdot V = 927.4 \times 3.054 \times 10^{-12} \approx 2.832 \times 10^{-9} \, \text{kg}\,.
\]
For 2 kg of microspheres, the number of microspheres is:
\[
N = \frac{2}{2.832 \times 10^{-9}} \approx 7.063 \times 10^8.
\]
For 200 g of MeSA, the number of moles is:
\[
n = \frac{200}{152.149} \approx 1.314 \, \text{mol}\,.
\]
Using Avogadro’s number ($ N_A = 6.022 \times 10^{23} \, \text{molecules}~\si{\per\mol} $), the number of MeSA molecules, corresponding to $200~\si{\gram}$ of MeSA is
\[
N_{\text{molecules}} = 1.314 \times 6.022 \times 10^{23} \approx 7.913 \times 10^{23}.
\]
Finally, using $N$ and $N_{\text{molecules}}$, the average number of MeSA molecules per microsphere can be computed as follows
\[
\bar{N} = \frac{N_{\text{molecules}}}{N} = \frac{7.913 \times 10^{23}}{7.063 \times 10^8} \approx 1.121 \times 10^{15}.
\]

\ifCLASSOPTIONcaptionsoff
  \newpage
\fi
\bibliographystyle{IEEEtran}
\bibliography{Bibliography}

\begin{thebibliography}{10}
\providecommand{\url}[1]{#1}
\csname url@samestyle\endcsname
\providecommand{\newblock}{\relax}
\providecommand{\bibinfo}[2]{#2}
\providecommand{\BIBentrySTDinterwordspacing}{\spaceskip=0pt\relax}
\providecommand{\BIBentryALTinterwordstretchfactor}{4}
\providecommand{\BIBentryALTinterwordspacing}{\spaceskip=\fontdimen2\font plus
\BIBentryALTinterwordstretchfactor\fontdimen3\font minus \fontdimen4\font\relax}
\providecommand{\BIBforeignlanguage}[2]{{%
\expandafter\ifx\csname l@#1\endcsname\relax
\typeout{** WARNING: IEEEtran.bst: No hyphenation pattern has been}%
\typeout{** loaded for the language `#1'. Using the pattern for}%
\typeout{** the default language instead.}%
\else
\language=\csname l@#1\endcsname
\fi
#2}}
\providecommand{\BIBdecl}{\relax}
\BIBdecl

\bibitem{Akyildiz2008}
I.~F. Akyildiz, F.~Brunetti, and C.~Blázquez, ``Nanonetworks: A new communication paradigm,'' \emph{Comp. Netw.}, vol.~52, no.~12, pp. 2260--2279, 2008.

\bibitem{Nakano2013}
T.~Nakano, A.~W. Eckford, and T.~Haraguchi, \emph{Molecular Communication}.\hskip 1em plus 0.5em minus 0.4em\relax Cambridge University Press, 2013.

\bibitem{9488662}
H.~K. Rudsari \emph{et~al.}, ``Targeted drug delivery for cardiovascular disease: Modeling of modulated extracellular vesicle release rates,'' \emph{IEEE Trans, NanoBiosci.}, vol.~20, no.~4, pp. 444--454, 2021.

\bibitem{Aktas_odor}
D.~Aktas, B.~E. Ortlek, M.~Civas, E.~Baradari, A.~B. Kilic, F.~E. Bilgen, A.~S. Okcu, M.~Whitfield, O.~Cetinkaya, and O.~B. Akan, ``Odor-based molecular communications: State-of-the-art, vision, challenges, and frontier directions,'' \emph{IEEE Commun. Surv. Tuts.}, pp. 1--34, 2024.

\bibitem{expMC_Seb}
S.~Lotter, L.~Brand, V.~Jamali, M.~Schäfer, H.~M. Loos, H.~Unterweger, S.~Greiner, J.~Kirchner, C.~Alexiou, D.~Drummer, G.~Fischer, A.~Buettner, and R.~Schober, ``Experimental research in synthetic molecular communications – part ii,'' \emph{IEEE Nanotechnol. Mag.}, vol.~17, no.~3, pp. 54--65, 2023.

\bibitem{Felicetti_MC}
L.~Felicetti, M.~Femminella, and G.~Reali, ``Molecular communications in blood vessels: Models, analysis, and enabling technologies,'' \emph{Commun. ACM}, vol.~68, no.~3, p. 60–69, Feb. 2025.

\bibitem{vahid_Odor}
V.~Jamali \emph{et~al.}, ``Olfaction-inspired {MC}s: Molecule mixture shift keying and cross-reactive receptor arrays,'' \emph{IEEE Trans. Commun.}, vol.~71, no.~4, pp. 1894--1911, 2023.

\bibitem{Gould_WE}
F.~Gould, Z.~S. Brown, and J.~Kuzma, ``Wicked evolution: Can we address the sociobiological dilemma of pesticide resistance?'' \emph{Science}, vol. 360, no. 6390, pp. 728--732, 2018.

\bibitem{Goulson_OV}
D.~Goulson, ``Review: An overview of the environmental risks posed by neonicotinoid insecticides,'' \emph{J. Appl. Ecol.}, vol.~50, no.~4, pp. 977--987, 2013.

\bibitem{Stamati_Chem}
P.~Nicolopoulou-Stamati, S.~Maipas, C.~Kotampasi, P.~Stamatis, and L.~Hens, ``Chemical pesticides and human health: The urgent need for a new concept in agriculture,'' \emph{Fron. Public Health}, vol. Volume 4 - 2016, 2016.

\bibitem{Baldwin2006}
I.~T. Baldwin, R.~Halitschke, A.~Paschold, C.~C. von Dahl, and C.~A. Preston, ``Volatile signaling in plant-plant interactions: "talking trees" in the genomics era,'' \emph{Science}, vol. 311, no. 5762, pp. 812--815, 2006.

\bibitem{Heil2010}
M.~Heil and R.~Karban, ``Explaining evolution of plant communication by airborne signals,'' \emph{Trends Ecol. Evol.}, vol.~25, no.~3, pp. 137--144, 2010.

\bibitem{Turlings2018}
T.~C. Turlings and M.~Erb, ``Tritrophic interactions mediated by herbivore-induced plant volatiles: Mechanisms, ecological relevance, and application potential,'' \emph{Annual Rev. Entomol.}, vol.~63, no. Volume 63, 2018, pp. 433--452, 2018.

\bibitem{Hamdan_InCel}
H.~Awan \emph{et~al.}, ``Modeling the role of inter-cellular communication in modulating photosynthesis in plants,'' \emph{IEEE Trans. Mol. Biol. Multi-Scale Commun.}, vol.~7, no.~2, pp. 94--99, 2021.

\bibitem{Hamdan_ComPlant}
H.~Awan, K.~Zeid, R.~S. Adve, N.~Wallbridge, C.~Plummer, and A.~W. Eckford, ``Communication in plants: Comparison of multiple action potential and mechanosensitive signals with experiments,'' \emph{IEEE Trans. NanoBiosci.}, vol.~19, no.~2, pp. 213--223, 2020.

\bibitem{Hamdan_InfTheo}
H.~Awan \emph{et~al.}, ``Information theoretic based comparative analysis of different communication signals in plants,'' \emph{IEEE Access}, vol.~7, pp. 117\,075--117\,087, 2019.

\bibitem{Imen_Plant}
I.~Bekkari, S.~Lombardo, A.~Coviello, and M.~Magarini, ``Detecting severe plant water stress with machine learning in iot-enabled chamber,'' in \emph{7th Int. Balk. Conf. Commun. Netw.}, 2024, pp. 153--157.

\bibitem{Hughes_plant}
A.~Hughes, C.~Faulkner, R.~J. Morris, and M.~Tomkins, ``Intercellular communication as a series of narrow escape problems,'' \emph{IEEE Trans. Mol. Biol. Multi-Scale Commun.}, vol.~7, no.~2, pp. 89--93, 2021.

\bibitem{Unluturk_ETE}
B.~D. Unluturk and I.~F. Akyildiz, ``An end-to-end model of plant pheromone channel for long range molecular communication,'' \emph{IEEE Trans. NanoBiosci.}, vol.~16, no.~1, pp. 11--20, 2017.

\bibitem{Bilgen:fungi:2024}
F.~E. Bilgen and O.~B. Akan, ``Mycorrhizal fungi and plant symbiosis for energy harvesting in the internet of plants,'' in \emph{Proc. 11th ACM Int. Conf. Nanosc. Comp. Commun.}, ser. NANOCOM '24.\hskip 1em plus 0.5em minus 0.4em\relax New York, NY, USA: Association for Computing Machinery, 2024, p. 35–40.

\bibitem{su17052018}
R.~Durak, M.~Materowska, R.~Hadley, L.~Oosterhuis, T.~Durak, and B.~Borowiak-Sobkowiak, ``The role of flower strips in increasing beneficial insect biodiversity and pest control in vineyards,'' \emph{Sustainability}, vol.~17, 2025.

\bibitem{KORSMEYER1981211}
R.~W. Korsmeyer and N.~A. Peppas, ``Effect of the morphology of hydrophilic polymeric matrices on the diffusion and release of water soluble drugs,'' \emph{J. Membr. Sci.}, vol.~9, no.~3, pp. 211--227, 1981.

\bibitem{Ryu2003}
C.-M. Ryu, M.~A. Farag, C.-H. Hu, M.~S. Reddy, H.-X. Wei, P.~W. Paré, and J.~W. Kloepper, ``Bacterial volatiles promote growth in {A}rabidopsis,'' \emph{Proc. Nat. Acad. Sci.}, vol. 100, no.~8, pp. 4927--4932, 2003.

\bibitem{Olde2014}
G.~E.~D. Oldroyd, ``Speak, friend, and enter: signalling systems that promote beneficial symbiotic associations in plants,'' \emph{Nat. Rev. Microbiol.}, vol.~11, no.~4, pp. 252--263, 4 2013.

\bibitem{Dewhirst2010}
S.~Y. Dewhirst, J.~A. Pickett, and J.~Hardie, ``Aphid pheromones,'' \emph{Vitamins and Hormones}, vol.~83, pp. 551--574, 2010.

\bibitem{Rowen2017}
E.~Rowen, M.~Gutensohn, N.~Dudareva, and I.~Kaplan, ``Carnivore attractant or plant elicitor? {M}ultifunctional roles of methyl salicylate lures in tomato defense,'' \emph{J. Chem. Ecol.}, vol.~43, no.~6, pp. 573--585, 2017.

\bibitem{Yi2016}
H.-S. Yi, Y.-R. Ahn, G.~C. Song, S.-Y. Ghim, S.~Lee, G.~Lee, and C.-M. Ryu, ``Impact of a bacterial volatile 2,3-butanediol on bacillus subtilis rhizosphere robustness,'' \emph{Front. Microbiol.}, vol.~7, p. 993, 2016.

\bibitem{chromatography}
L.~Cai, J.~A. Koziel, and M.~E. O'Neal, ``Studying plant–insect interactions with solid phase microextraction: Screening for airborne volatile emissions response of soybeans to the soybean aphid, aphis glycines matsumura (hemiptera: Aphididae),'' \emph{Chromatography}, vol.~2, no.~2, pp. 265--276, 2015.

\bibitem{Fontana2011}
A.~Fontana, M.~Held, C.~A. Fantaye, T.~C. Turlings, J.~Degenhardt, and J.~Gershenzon, ``Attractiveness of constitutive and herbivore-induced sesquiterpene blends of maize to the parasitic wasp {Cotesia marginiventris} (cresson),'' \emph{J. Chem. Ecol.}, vol.~37, no.~6, pp. 582--591, 6 2011.

\bibitem{Jianwei}
J.~Su, S.~Zhu, Z.~Zhang, and F.~Ge, ``{Effect of Synthetic Aphid Alarm Pheromone (E)-$\beta$-Farnesene on Development and Reproduction of Aphis gossypii (Homoptera: Aphididae)},'' \emph{J. Econ. Entomol.}, vol.~99, no.~5, pp. 1636 -- 1640, 2006.

\bibitem{Maillet2011}
F.~Maillet, V.~Poinsot, O.~André, V.~Puech-Pagès, A.~Haouy, M.~Gueunier, L.~Cromer, D.~Giraudet, D.~Formey, A.~Niebel, E.~A. Martinez, H.~Driguez, G.~Bécard, and J.~Dénarié, ``Fungal lipochitooligosaccharide symbiotic signals in arbuscular mycorrhiza,'' \emph{Nature}, vol. 469, no. 7328, pp. 58--63, 1 2011.

\bibitem{wmo2018guide}
{World Meteorological Organization}, \emph{Guide to Meteorological Instruments and Methods of Observation (WMO-No. 8)}, 2018th~ed., World Meteorological Organization, Geneva, Switzerland, 2018.

\bibitem{Stull1988}
R.~B. Stull, \emph{An Introduction to Boundary Layer Meteorology}, 1st~ed., ser. Atmospheric and Oceanographic Sciences Library.\hskip 1em plus 0.5em minus 0.4em\relax Springer Dordrecht, 1988.

\bibitem{FDM_Book}
R.~J. LeVeque, \emph{Finite difference methods for ordinary and partial differential equations: steady-state and time-dependent problems}.\hskip 1em plus 0.5em minus 0.4em\relax SIAM, 2007.

\bibitem{Courant1928}
R.~Courant \emph{et~al.}, ``{\"U}ber die partiellen {D}ifferenzengleichungen der mathematischen {P}hysik,'' \emph{Math. Annalen}, vol. 100, no.~1, pp. 32--74, 1928.

\bibitem{ephrath1996modelling}
J.~Ephrath, J.~Goudriaan, and A.~Marani, ``Modelling diurnal patterns of air temperature, radiation wind speed and relative humidity by equations from daily characteristics,'' \emph{Agricultural systems}, vol.~51, no.~4, pp. 377--393, 1996.

\bibitem{LOUKATOU20181328}
A.~Loukatou, S.~Howell, P.~Johnson, and P.~Duck, ``Stochastic wind speed modelling for estimation of expected wind power output,'' \emph{Applied Energy}, vol. 228, pp. 1328--1340, 2018.

\bibitem{Daikoku2007}
K.~Daikoku, S.~Hattori, A.~Deguchi, Y.~Fujita, H.~Park, and K.~Matsumoto, ``Impact of wind direction on diurnal and seasonal changes in wind profiles,'' \emph{J. Forest Research}, vol.~12, no.~6, pp. 452--466, 12 2007.

\bibitem{Bird2002}
R.~B. Bird, W.~E. Stewart, and E.~N. Lightfoot, \emph{Transport Phenomena}, 2nd~ed.\hskip 1em plus 0.5em minus 0.4em\relax New York: John Wiley \& Sons, Inc., 2002.

\end{thebibliography}

\end{document}